\documentclass[prl,twocolumn,a4paper, superscriptaddress,showpacs]{revtex4}
\usepackage{graphicx,amsmath,bm,amsfonts,color}

\definecolor{nred}{rgb}{0.7,0.2,0.2}
\definecolor{nblack}{rgb}{0,0,0}
\definecolor{nblue}{rgb}{0.2,0.2,0.7}
\definecolor{ngreen}{rgb}{0.2,0.6,0.2}

\begin{document}

\title{Witnessing trustworthy single-photon entanglement \\
with local homodyne measurements}
\date{\today}
\author{Olivier Morin\footnotemark[1] \footnotetext{\footnotemark[1] These authors contributed equally
to this work.}}
\affiliation{Laboratoire Kastler Brossel, Universit\'e Pierre et Marie Curie, Ecole Normale Sup\'erieure, CNRS, 4 place Jussieu, 75252 Paris Cedex 05, France}
\author{Jean-Daniel Bancal\footnotemark[1]}
\affiliation{Group of Applied Physics, University of Geneva, CH-1211 Geneva 4, Switzerland}
\author{Melvyn Ho}
\affiliation{Centre for Quantum Technologies, National University of Singapore, 3 Science Drive 2, Singapore 117543}
\author{Pavel Sekatski}
\affiliation{Group of Applied Physics, University of Geneva, CH-1211 Geneva 4, Switzerland}
\author{Virginia D'Auria}
\affiliation{Laboratoire de Physique de la Mati\`ere Condens\'ee, CNRS UMR 7336, Universit\'e de Nice - Sophia Antipolis, Parc Valrose, 06108 Nice Cedex 2, France}
\author{ \\ Nicolas Gisin}
\affiliation{Group of Applied Physics, University of Geneva, CH-1211 Geneva 4, Switzerland}
\author{Julien Laurat}
\affiliation{Laboratoire Kastler Brossel, Universit\'e Pierre et Marie Curie, Ecole Normale Sup\'erieure, CNRS, 4 place Jussieu, 75252 Paris Cedex 05, France}
\author{Nicolas Sangouard}
\affiliation{Group of Applied Physics, University of Geneva, CH-1211 Geneva 4, Switzerland}

\begin{abstract}
Single-photon entangled states, i.e. states describing two optical paths sharing a single photon, constitute the simplest form of entanglement. Yet they provide a valuable resource in quantum information science. Specifically, they lie at the heart  of quantum networks, as they can be used for quantum teleportation, swapped and purified with linear optics. The main drawback of such entanglement is the difficulty in measuring it. Here, we present and experimentally test an entanglement witness allowing one not only to say whether a given state is path-entangled but also that entanglement lies in the subspace where the optical paths are each filled with one photon at most, i.e. refers to single-photon entanglement. It uses local homodyning only and relies on no assumption about the Hilbert space dimension of the measured system. Our work provides a simple and trustful method for verifying the proper functioning of future quantum networks.
\end{abstract}
\pacs{03.65.Ud, 03.67.Mn, 42.50.Dv}
\maketitle

\paragraph{Motivations}
Quantum networks \cite{Kimble08} provide broad capabilities, ranging from long distance quantum communication at large scales \cite{Briegel98, Duan01}, to the simulation of quantum many-body systems \cite{Illuminati06} in tabletop implementations. Remarkable progresses have been made in practice \cite{progress_networks1, progress_networks2, progress_networks3} and experimental capabilities are now advancing into a domain of rudimentary functionality for quantum nodes connected by quantum channels \cite{Chou07, Yuan08, Moehring07, Ritter12}. Surprisingly, the task of checking that a newly implemented quantum network performs well remains non-trivial. \\
In the past decade, a great number of architectures based on atomic ensembles and linear optics have been proposed \cite{Sangouard11}. We now know that  quantum networks based on single-photon entanglement \cite{vanEnk05}, i.e. entangled states of the form 
\begin{equation}
\label{single_ph_ent}
\frac{1}{\sqrt{2}}\left(|1\rangle_A|0\rangle_B+|0\rangle_A|1\rangle_B\right)
\end{equation}
where $A$ and $B$ are two spatial modes sharing a delocalized photon, are very attractive: They require significantly fewer resources than the other architectures and are less sensitive to memory and photon detector inefficiencies \cite{Sangouard11}. Furthermore, they are efficient when combined with temporal multiplexing \cite{Simonc07}.  However, such networks have a major drawback: The detection of single-photon entangled states is very challenging. One cannot resort, for example, to violating a Bell inequality given solely photon counting techniques. \\

Hitherto, there are three prescribed methods to detect single-photon entanglement. The first one converts two copies of a single-photon entangled state into one copy of two-particle entanglement. Starting from entanglement $(|1\rangle_{A_1}|0\rangle_{B_1}+|0\rangle_{A_1}|1\rangle_{B_1})\otimes(|1\rangle_{A_2}|0\rangle_{B_2}+|0\rangle_{A_2}|1\rangle_{B_2})$ between the modes $A_1$ and $B_1$ and between $A_2$ and $B_2,$ it basically consists of a post-selective projection onto the subspace with one excitation in each location, yielding $|1\rangle_{A_1}|1\rangle_{B_2}+|1\rangle_{A_2}|1\rangle_{B_1}$ \cite{Duan01}. The latter is analogous to conventional polarization or time-bin entanglement
and any witness suited for such entanglement can thus be used to post-selectively detect single-photon entanglement. Nevertheless, this approach is not fully satisfying conceptually because it relies on post-selection. Furthermore, for practical implementation, the need to create two copies requires twice the number of resources at each node.\\ 
The second method is based on partial quantum state tomography. Specifically, one reconstructs a reduced density matrix that corresponds to a projection of the full density matrix into a subspace with at most one photon locally. The presence of entanglement is then inferred from an entanglement measure computed from the reduced density matrix \cite{Chou05}. Specifically, this tomographic approach requires the knowledge of probabilities $p_{mn}$ of having $m$ photons in mode A and $n$ in mode B, where $m,n \in \{0,1\},$ and the visibility $V$ of the single-photon interference pattern obtained by combining the modes A \& B into a beam-splitter. Although it has triggered highly successful experiments \cite{Chou05, Choi08, Usmani11, Lee11}, the approach presented in Ref. \cite{Chou05} cannot be directly used in large scale networks when one needs to check the entanglement between far away locations, since the knowledge of $V$ relies on a joint measurement of A \& B modes. \\
The last method uses local homodyne detections and provides, a priori, a full tomography of the state that can subsequently be used to measure the entanglement \cite{Babichev04,Note_Babichev04}. However, 
the tomographic approach requires a number of measurements which increases with the dimension of the state being measured \cite{Leonhardt96}. In practice, one could be tempted to make an assumption on the regularity of the measured Wigner function to reduce the number of measurements or, equivalently, on the dimension of the system's Hilbert space, especially when focusing on single-photon entanglement. But this would amount to make an assumption about the system that we want to characterize. 
One can also estimate the dimension of the state from measurements but it is not clearly established how errors on this estimation can affect the conclusion about the presence of entanglement. More generally, the exponential increase of required measurements with the number of measured subsystems makes the tomography not suited to decide on the presence of entanglement in quantum networks \cite{Comment_tomo}, contrary to entanglement witnesses \cite{Guhne09}.\\

\begin{figure}[ht!]
\includegraphics[width=7 cm]{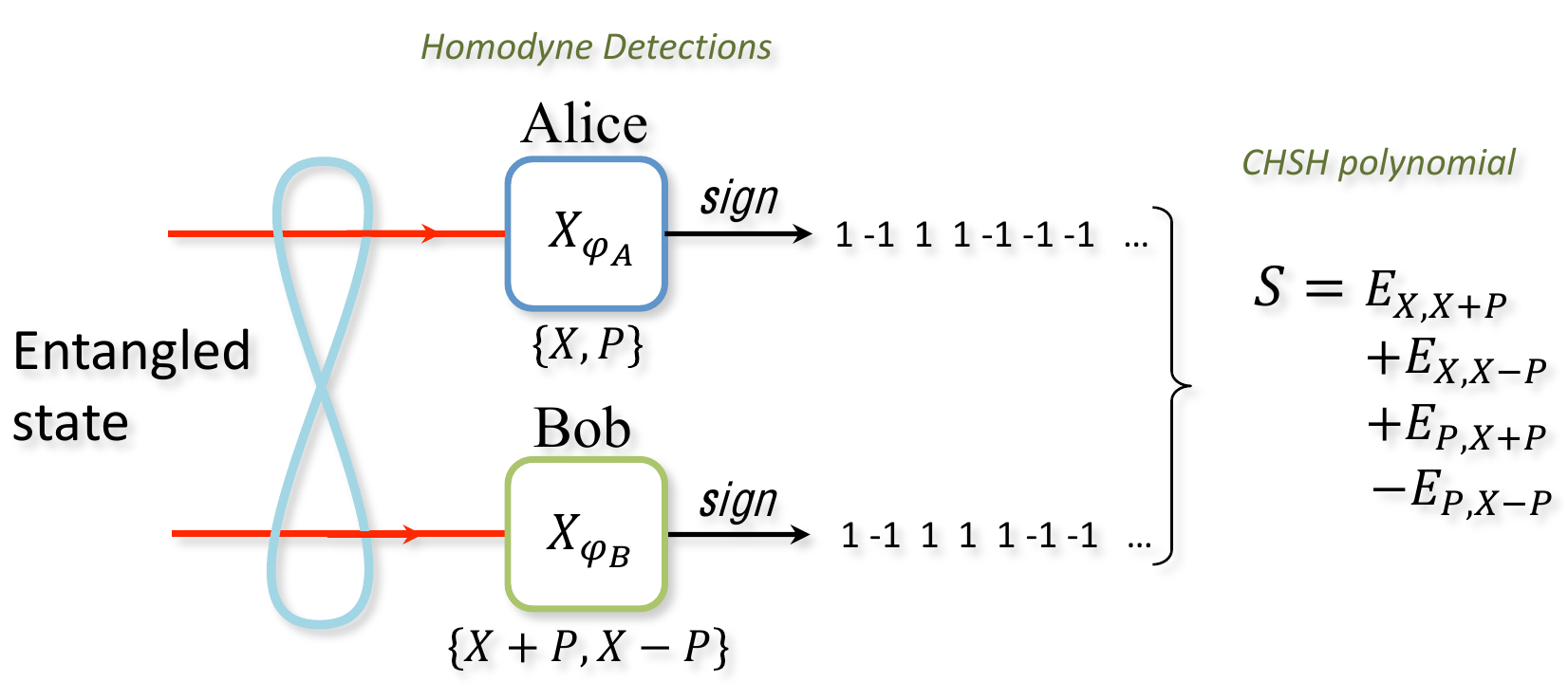}
\caption{Principle of the proposed entanglement witness.}
\label{fig1}
\end{figure}

\paragraph{Principle} Here, we propose a simple approach to witness single-photon entanglement which relies on local measurements only and needs neither post-selection nor assumption on the tested state. The basic principle is drawn in Fig. \ref{fig1}. 
Two distant observers, Alice \& Bob, share a quantum state. To check whether it is entangled, each of them randomly chooses a measurement among two quadratures, $\{ X, P\}$ for Alice and $\{X+ P, X- P\}$ for Bob. At each run, they obtain a real number. They then process the results to get binary outcomes using a sign binning, i.e. they attribute the result $-1$ if the result is negative and $+1$ otherwise. By repeating the experiment several times, Alice \& Bob can compute the conditional probabilities $p(a,b|x,y)$ where $a,b \in \{-1,+1\},$ $x \in \{ X,  P\}$ and $y \in \{ X+ P, X- P\}.$ Substituting these probabilities by their values into the Clauser-Horne-Shimony-Holt (CHSH) \cite{Clauser69} polynomial
\begin{equation}
\label{CHSH}
S=E_{X,X+P}+E_{X,X-P}+E_{P,X+P}-E_{P,X-P},
\end{equation}
where $E_{x,y}=\sum_{a,b=\{-1,+1\}} p(a=b|x,y)-p(a \neq b|x,y),$ they obtain a real number $S.$\\
The value of $S$ can easily be obtained under the assumption that Alice \& Bob have each a qubit. Indeed, in the Fock basis $\{|0\rangle,|1\rangle\},$ the measurement of the $X$ quadrature with sign binning is equivalent to a noisy $\sigma_x$ measurement \cite{Quintino11, Sangouard11bis}, 
$
\int_{-\infty}^{0} dx \-\ |x\rangle\langle x|-\int_{0}^{\infty} dx \-\ |x\rangle\langle x|= \sqrt{\frac{2}{\pi}} \sigma_x
$
and similarly, the $P$ quadrature corresponds to $\sigma_y$ with the same noise. For the setting choice $\{X, P\}$ and $\{X+P, X-P\},$ the state  (\ref{single_ph_ent}) thus yields to  $S=2\sqrt{2}�\times \frac{2}{\pi} \approx 1.8.$ Furthermore, the maximum value that can be obtained with a separable state belonging to the subspace $\{|0\rangle, |1\rangle\}^{\otimes 2}$ is $S_{\rm{sep}} = \sqrt{2} \times \frac{2}{\pi} \approx 0.9$ \cite{Roy05}. Since $S$ is smaller than $2,$ the proposed CHSH-like test does not highlight the non-local characteristic of a single photon delocalized among two modes, but it does provide an attractive entanglement witness: If the measured CHSH value is larger than $S_{\rm{sep}},$ Alice \& Bob can conclude that they share an entangled state.\\
This holds for qubits only. In practice, however, the state describing the modes A \& B includes multi-photon components, and does not reduce to a two-qubit state. We show below how the entanglement witness can be extended to the case of arbitrary dimensional bipartite states. First, we show how Alice \& Bob can accurately estimate the probability that their state lies out of a two-qubit space $\{|0\rangle, |1\rangle\}^{\otimes 2}.$ We then demonstrate that this probability can be used to upper bound the maximal CHSH value that can be obtained with separable states. \\

\paragraph{Bounding the Hilbert space dimension} Let us consider the case where Alice \& Bob do not have qubits, but quantum states of arbitrary dimension. First, they need to bound the probability that at least one of their modes is populated with more than one photon $p(n_A \geq 2 \cup n_B \geq 2).$ This can be realized without assumption on the Hilbert space dimension, by first determining the probabilities $p(n_A=j)$ $(p(n_B=j))$ of having $j$ photons in Alice's (Bob's) mode using local homodyning with phase averaged local oscillators through a direct integration of the obtained data with a pattern function \cite{Munroe95}. The joint probability $p(n_A \geq 2 \cup n_B \geq 2)=p(n_A \geq 2)+ p(n_B \geq 2)-p(n_A \geq 2 \cap n_B \geq 2)$ can then be bounded by the parameter $p^\star$ defined as follows
\begin{equation}
\label{p_star}
p^\star=2-\left(\sum_{j=0}^{1} p\left(n_A=j\right)+p\left(n_B=j\right)\right).
\end{equation}
We now show how the knowledge of $p^\star$ can be used to construct an operational witness for single-photon entanglement.\\

\paragraph{Evaluating the maximal CHSH value with separable states}
\begin{figure}[ht!]
\includegraphics[width=7 cm]{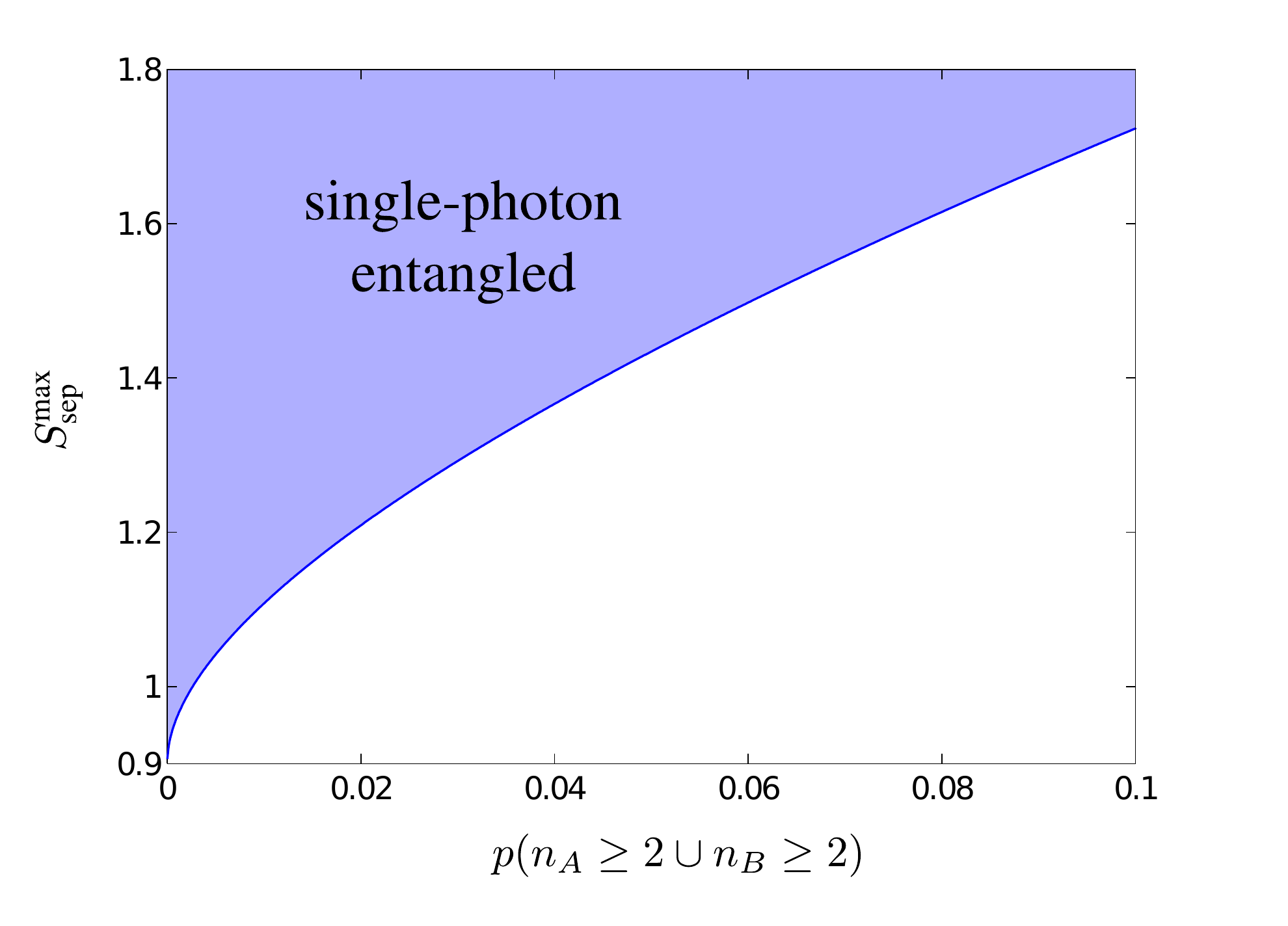}
\caption{Separable bound $S^{\max}_{\rm sep}$ as a function of the probability that at least one of the two protagonists gets more than one photon $p(n_A \geq 2 \cup n_B \geq 2).$ The blue curve allows one to know the maximum value of the CHSH polynomial $S^{\max}_{\rm sep}$ that a state separable in the $\{|0\rangle,|1\rangle\}^{\otimes 2}$ subspace can reach. If $S_{\rm obs} > S^{\max}_{\rm sep},$ one can conclude that the projection of the state in the subspace with zero and one photon locally is entangled.}
\label{fig2}
\end{figure}
Consider the general case where $p^\star \neq 0,$ i.e. $p(n_A \geq 2 \cup n_B \geq 2) \neq 0$ a priori. The state of Alice \& Bob can be described by the density matrix 
\begin{equation}
\label{share_state}
\rho=\begin{pmatrix}
  \rho_{n_A \leq 1 \cap n_B \leq 1}    &  \rho_{\rm{coh}}  \\
    \rho_{\rm{coh}}^\dag  &  \rho_{n_A \geq 2 \cup n_B \geq 2} 
\end{pmatrix}
\end{equation}
where $\rho_{n_A \leq 1 \cap n_B \leq 1}$ denotes the $4 \times 4$ block with at most one photon per mode, $\rho_{n_A \geq 2 \cup n_B \geq 2}$ refers to the block where at least one of the two modes contains at least two photons and $\rho_{\rm{coh}}$ is associated to the coherence between these two blocks. Since $\rho_{n_A \geq 2 \cup n_B \geq 2}$ possibly spans a Hilbert space of infinite dimension, there could be an infinite number of coherence terms. However, 
a few of them give a non-zero contribution to the CHSH polynomial if a phase averaged homodyne detection is used at each location. Specifically, consider the case where Alice \& Bob perform the measurements $ X_{\varphi_A} = \cos \varphi_A X +  \sin \varphi_A P$ and $X_{\varphi_B} = \cos \varphi_B X +  \sin \varphi_B P$ respectively, where $\varphi_A$ and $\varphi_B$ are random variables such that $\langle e^{i k \varphi_{A,B}} \rangle=0,$ $k \in \mathbb{N}^{\star}$ but where the phase difference $\varphi_A-\varphi_B=\Delta \varphi$ is fixed. This only requires classical but not quantum communication, and hence can only decrease the entanglement that Alice \& Bob potentially share. In particular, if Alice can choose a measurement among the two quadatures $\{X_{\varphi_A^1}, X_{\varphi_A^2}\}$ and if Bob's choice reduces to one of the quadratures $\{X_{\varphi_B^1}, X_{\varphi_B^2}\}$ such that $\varphi_A^1-\varphi_B^1=-\frac{\pi}{4},$ $\varphi_A^1-\varphi_B^2=\frac{\pi}{4},$ $\varphi_A^2-\varphi_B^1=\frac{\pi}{4}$ and $\varphi_A^2-\varphi_B^2=\frac{3\pi}{4},$ we show in the appendix that the CHSH polynomial corresponding to the state (\ref{share_state}) is bounded by
\begin{eqnarray}
\label{S_averaged}
&S^{\max}& =\frac{16}{\sqrt{2} \pi} \Re\Big[\langle 01 | \rho_{n_A \leq 1 \cap n_B \leq 1} | 10\rangle\Big]\\
\nonumber
&& + \frac{8}{\pi} \left(\Re\Big[\langle 20 | \rho_{\rm{coh}} | 11\rangle\Big]+\Re\Big[\langle 02 | \rho_{\rm{coh}} | 11\rangle\Big]  \right)\\
\nonumber
&& +2\sqrt{2} p(n_A \geq 2 \cup n_B \geq 2)
\end{eqnarray}
where $\Re$ denotes the real part. For a given value of $p(n_A \geq 2 \cup n_B \geq 2),$ $S^{\max}$ can be directly maximized over the set of physical states $(\rho \in \{|0\rangle,|1\rangle,|2\rangle\}^{\otimes 2}, \text{tr}(\rho) \leq 1,\rho \geq 0)$ that satisfy the observed photon number distributions, i.e. $p_{00}+p_{10}+p_{01}+p_{11}=1-p(n_A \geq 2 \cup n_B \geq 2)$ and that are separable in the $\{|0\rangle,|1\rangle\}^{\otimes 2}$ subspace, i.e. for which the projection into this subspace remains positive under partial transposition (PPT) \cite{Peres96, Horodecki96}. Figure \ref{fig2} shows the result of this optimization $S^{\max}_{\rm sep}.$\\

\paragraph{Witnessing single-photon entanglement} 
This provides a truly state-independent witness \cite{explain} of entanglement: First, the protagonists determine the local photon-number distributions, from which they deduce an upper bound $p^\star$ on the joint probability $p(n_A \geq 2 \cup n_B \geq 2).$ Secondly, they deduce $S^{\max}_{\rm sep}(p^\star)$ form Fig. \ref{fig2} (See the appendix). Thirdly, they measure the CHSH value $S_{\text obs}$ by randomly choosing measurements among $\{X_{\varphi_A^1}, X_{\varphi_A^2}\}$ and $\{X_{\varphi_B^1}, X_{\varphi_B^2}\}$ respectively and by subsequently computing the CHSH polynomial through Eq. (\ref{CHSH}). If $S_{\rm obs} > S^{\max}_{\rm sep}(p{^\star}),$ Alice \& Bob know that the projection of their state into the subspace $\{|0\rangle, |1\rangle\}^{\otimes 2}$ has a negative partial transpose, i.e. they can safely conclude that the state is entangled and that the entanglement resides in the subspace with at most one photon locally.\\
Importantly, a tighter bound can be obtained if $S^{\max}$ is maximized over the set of states with a positive partial transpose not only satisfying $p_{00}+p_{10}+p_{01}+p_{11}=1-p(n_A \geq 2 \cup n_B \geq 2)$ but also reproducing the locally measured probabilities $p(n_A=j)$ $(p(n_B=j))$ for having $j={0,1}$ photon in Alice's (Bob's) mode. These additional constraints have been taken into account for the computation of the separable bounds related to the experiment presented below (see appendix).  \\

\begin{figure}[htb]
\centerline{\includegraphics[width=0.4\textwidth]{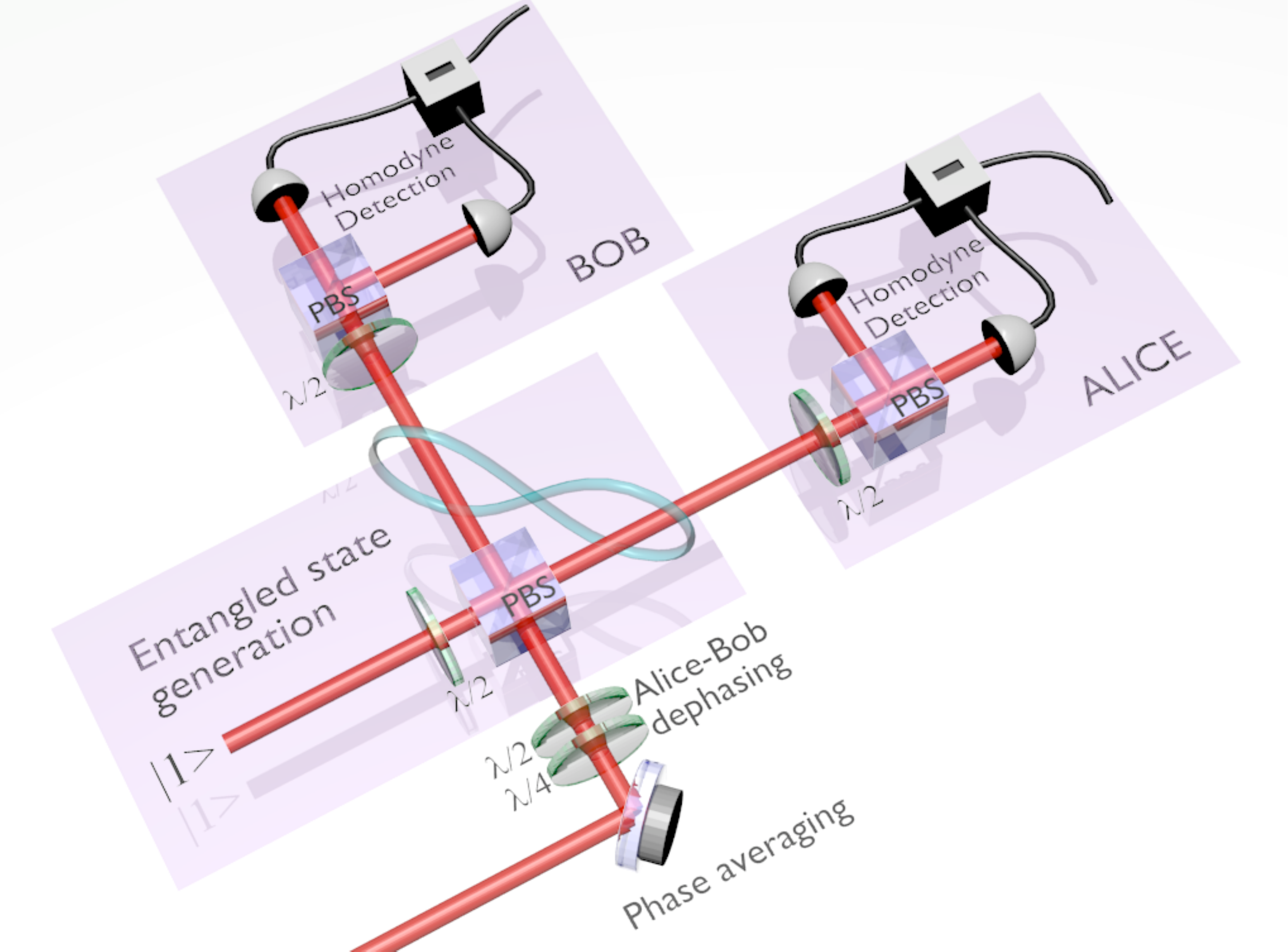}}
\caption{Experimental setup. A tunable single-photon entangled state is created by sending a heralded single photon on a tunable beam-splitter based on a polarizing beam-splitter (PBS) and a half-wave plate ($\lambda/2$). The proposed witness is then tested with two independent homodyne detections (Alice \& Bob). The local oscillator is superposed to each modes via the first PBS. Its global phase is swept with a piezoelectric actuator. The relative phase $\Delta\varphi$ is set with a combination of birefringent plates.}
\label{fig3}
\end{figure}

\begin{figure}[ht!]
\includegraphics[width=6.5 cm]{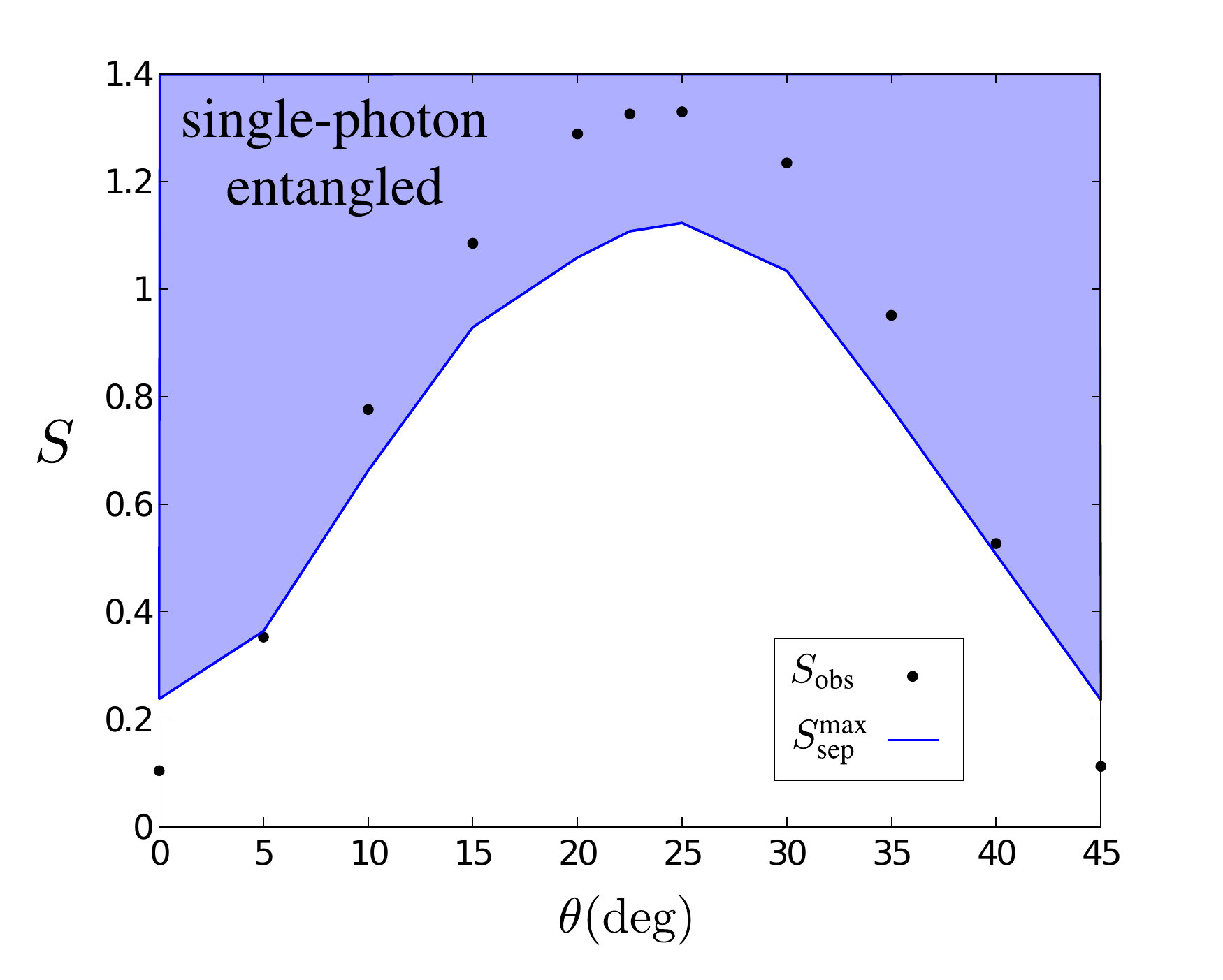}
\caption{Observed CHSH values $S_{\rm obs}$ (the size of points accounts for statistical errors) and separable bound $S_{\rm sep}^{\max}$ as a function of beam-splitter angle $\theta$. When $S_{\rm obs}>S_{\rm sep}^{\max}$, one can conclude that the measured state is entangled.}
\label{fig4}
\end{figure}

\paragraph{Proof-of-principle experiment} We start off with a heralded single photon generated by a conditional preparation technique operated on a two-mode squeezed vacuum emitted by a type-II optical parametric oscillator \cite{Morin12}. Without correction for detection loss, the overall fidelity reaches 70\%. Single-photon entanglement is obtained by sending the created photon into a beam-splitter \cite{Asboth05}. Specifically, by controlling the angle $\theta$ of a  half-wave plate relative to the axis of a polarizing beam-splitter (PBS), we create a tunable single-photon entangled state $\cos(2\theta)|0\rangle_A|1\rangle_B + \sin(2\theta)|1\rangle_A |0\rangle_B$ between the two output modes of the PBS, as sketched on Fig. \ref{fig3}.\\
The local oscillators that Alice \& Bob need to reveal entanglement are obtained by impinging a bright beam on the second input of the PBS: one polarization mode is used as the quantum channel, the orthogonal one conveys the local oscillator. It is worth mentioning that this technique can be implemented over long distances as realized in field implementation of quantum key distribution \cite{Fossier09}. The relative phase between Alice's and Bob's local oscillators $\Delta\varphi$ is fixed by choosing an appropriate elliptical polarization of the bright beam just before the PBS  \cite{Laurat05}. In practice, the setting difference is calibrated by observing the dephasing of interference fringes (the quantum state is replaced here by a coherent state). A global phase averaging is obtained by sweeping a piezoelectric transducer located on the path of the bright beam before the PBS. \\
For each heralding event, Alice \& Bob each obtain a real valued outcome which is extracted from homodyne photocurrents. Accumulating 200000 events for each quadrature relative measurements, they deduce the value of the CHSH polynomial $S_{\rm obs}.$ The same homodyne measurements also provide the local photon number distributions which are used to compute the separable bound $S_{\rm sep}^{\max}.$ We emphasize that the separable bound is here obtained by maximizing the CHSH value over the set of separable states that fulfill the locally measured photon-number occupation probabilities $p(n_A=j)$ and $p(n_B=j)$ for having $j={0,1}$ photon in Alice \& Bob's mode respectively. Furthermore, it takes several errors into account, for example, errors related to the quadrature measurement imperfections were considered (see appendix). The procedure is repeated for various angles $\theta$ ranging from 0  to 45$^{\circ}.$ Fig. \ref{fig4} shows the main result, i.e. the observed CHSH values and the separable bounds as a function of $\theta.$ One sees that they both reach maximal values around $\theta=22.5^{\circ}$ where Alice \& Bob ideally share a maximally entangled state. The small deviation between the observed value $S_{\rm obs}(\theta=22.5^{\circ}) \approx 1.33$ and the CHSH value that would be obtained with a maximally entangled state $(1.8)$ demonstrates that the overall source and detection efficiencies are very high. Furthermore, the observed CHSH values are almost all larger than the separable bounds when dealing with entangled states ($\theta \neq 0^{\circ}, \-\ 45^{\circ}$). This shows the great robustness of the proposed witness. 

\paragraph{Conclusion} We have presented and experimentally tested a witness for single-photon entanglement which does not need postselection, uses local measurements only and does not rely on assumptions about the dimension of the measured state. Note that our witness can be easily adapted to detect few-photon entanglement without additional complications. We believe that it will naturally find applications in long distance quantum communication, allowing users to check whether two remote nodes of a given quantum network are entangled. One important challenge in this context is to reveal the entanglement shared by a large number of parties. Finding Bell inequalities that could be used as witnesses for multi-partite single-photon entanglement is work for future.\\

\paragraph{Acknowledgments} We thank Mikael Afzelius, Cyril Branciard, F\'elix Bussi\`eres, Claude Fabre, Ci Wen Lim, Val\'erio Scarani, Rob Thew and Nuala Timoney for helpful discussions. We acknowledge support by the ERA-NET CHIST-ERA under the QScale project, by the EU project Qessence, from the Swiss NCCR-QSIT and from the National Research Fund and the Ministry of Education, Singapore. Julien Laurat is a member of the Institut Universitaire de France.

\newpage

\appendix
\section{Apendix A: Witnessing single-photon entanglement in qudit spaces: Theory}
As mentioned in the main text, if Alice \& Bob can guarantee that they perform measurements on qubits, they can demonstrate entanglement by simply performing a Bell-like test by randomly choosing a setting among two quadratures, $\{ X, P\}$ for Alice and $\{X+P, X-P\}$ for Bob. If the resulting CHSH value is larger than 0.9, they can safely conclude that their state is entangled. In practice, however, it is challenging to show that the systems one is measuring are well described by qubits. Below, we detail the procedure to follow in the general case where Alice \& Bob have qudits.
\subsection{CHSH polynomial for phase averaged homodyning}
Let 
$$
\rho= \sum_{ijkl} c_{ijkl} |ij\rangle\langle kl|
 $$
be the state that Alice \& Bob share, $|ij\rangle$ describing the state with $i$ photons in Alice's mode and $j$ photons in Bob's one. Consider that the phase of local oscillators that are required for homodyning, is averaged, i.e. Alice \& Bob perform the measurements
\begin{eqnarray}
\nonumber
&&X_{\varphi_A^{\ell}} = \cos \varphi_A^{\ell} X +  \sin \varphi_A^{\ell} P,\\
\nonumber
&&X_{\varphi_B^{\bar \ell}} = \cos \varphi_B^{\bar \ell} X +  \sin \varphi_B^{\bar \ell} P
\end{eqnarray}
respectively, with $\varphi_A^{\ell}$ and $\varphi_B^{\bar \ell}$ random variables satisfying $\langle e^{i\varphi_{A}^{\ell}} \rangle=\langle e^{i\varphi_{B}^{\bar \ell}} \rangle=0.$ Further consider that the phase difference $\Delta \varphi^{\ell \bar \ell}=\varphi_A^{\ell}-\varphi_B^{\bar \ell}$ to be tunable such that Alice can choose a measurement among the two quadatures $\{X_{\varphi_A^1}, X_{\varphi_A^2}\}$ relative to Bob's choice $\{X_{\varphi_B^1}, X_{\varphi_B^2}\}.$ (This can only underestimate the entanglement in average because this can be realized by local operations and classical communications. Namely, they could each apply a random phase shift to a shared local oscillator such that the phase difference is fixed, subsequently choose a quadrature measurement locally and ignore the information about the individual phase shifts). At each run, Alice \& Bob obtain each a real number. They then process the results to get binary outcomes using a sign binning, i.e. they assign the result $-1$ if the real number is negative and $+1$ otherwise. By repeating this procedure many times, they can access the probability that both get $+1$ for instance, knowing that they chose $ X_{\varphi_A^{\ell}}$ and $X_{\varphi_B^{\bar \ell}}$
\begin{eqnarray}
\nonumber
&p(+1,+1|X_{\varphi_A^{\ell}},X_{\varphi_B^{\bar \ell}})&=  \sum_{ijkl} \langle e^{i \varphi_B^{\bar \ell} (i+j-(k+l))}\rangle c_{ijkl}  \\
\nonumber
&&\times e^{i\Delta\varphi^{\ell \bar \ell}(i-k)} \int_{0}^{\infty} dx \phi_i(x) \phi_k(x) \\
\nonumber
&&\times \int_{0}^{\infty} dy \phi_j(y) \phi_l(y) 
\end{eqnarray}    
where $\phi_i(x)=\langle x|i\rangle.$ One sees that the off-diagonal elements $ |ij\rangle\langle kl|$ with different numbers of photons $(i+j \neq k+l)$ do not contribute to the probabilities $p(a,b|X_{\varphi_A^{\ell}},X_{\varphi_B^{\bar \ell}}),$ with $a, b = \{-1, +1\}.$ Furthermore, since for all $n$ and $m$ having the same parity,
\begin{equation}
\nonumber
\int_{0}^{\infty} dy \phi_n(y) \phi_m(y)=\int_{-\infty}^{0} dy \phi_n(y) \phi_m(y) = \frac{1}{2}\delta_{n,m} \\
\end{equation}
the terms $c_{ijkl}$ for which either $i$-$k$ or $j$-$l$ is an odd number, have a zero contribution to the correlator
\begin{eqnarray}
\nonumber
&E_{X_{\varphi_A^{\ell}},X_{\varphi_B^{\bar \ell}}}=\sum_{a,b=\{-1,+1\}} & p(a=b|X_{\varphi_A^{\ell}},X_{\varphi_B^{\bar \ell}})\\
\nonumber
&&-p(a \neq b|X_{\varphi_A^{\ell}},X_{\varphi_B^{\bar \ell}}).
\end{eqnarray}
Finally, one easily checks that the remaining $c_{ijkl}$ terms satisfying
\begin{eqnarray}
\label{constraints1}
&& i+j = k+l,\\
\label{constraints2}
&& |i-k| =1 \mod 2,\\
\label{constraints3}
&& |j-l| =1 \mod 2,
\end{eqnarray} 
yield 
\begin{eqnarray}
&E_{ X_{\varphi_A^{\ell}},X_{\varphi_B^{\bar \ell}}}&= 4\sum_{ijkl} 
c_{ijkl}  \nonumber \\
 && \times\int_{0}^{\infty} dx \phi_i(x) \phi_k(x) e^{i\Delta\varphi^{\ell \bar \ell}(i-k)} \nonumber \\
 && \times\int_{0}^{\infty} dy \phi_j(y) \phi_l(y).  \nonumber
 \end{eqnarray}
Specifically, if $\Delta \varphi^{11}=-\frac{\pi}{4},$ $\Delta \varphi^{12}=\frac{\pi}{4},$ $\Delta \varphi^{21}=\frac{\pi}{4}$ and $\Delta \varphi^{22}=\frac{3\pi}{4},$ the value of the CHSH polynomial 
\begin{equation}
\nonumber
S= E_{X_{\varphi_A^1},X_{\varphi_B^1}}+E_{X_{\varphi_A^1},X_{\varphi_B^2}}+E_{X_{\varphi_A^2},X_{\varphi_B^1}}-E_{X_{\varphi_A^2},X_{\varphi_B^2}}
\end{equation}
is given by
\begin{eqnarray}
\label{CHSH_random}
\nonumber
&S=& 16  \sum_{ijkl}  c_{ijkl} \cos\left(\left(i-k\right)\frac{\pi}{4}\right)\\
&& \times \int_{0}^{\infty} dx \phi_i(x) \phi_k(x) \-\ . \int_{0}^{\infty} dy \phi_j(y) \phi_l(y).
\end{eqnarray}

\subsection{Optimizing the CHSH value over the set of separable states}
\begin{figure}[ht!]
\includegraphics[width=8 cm]{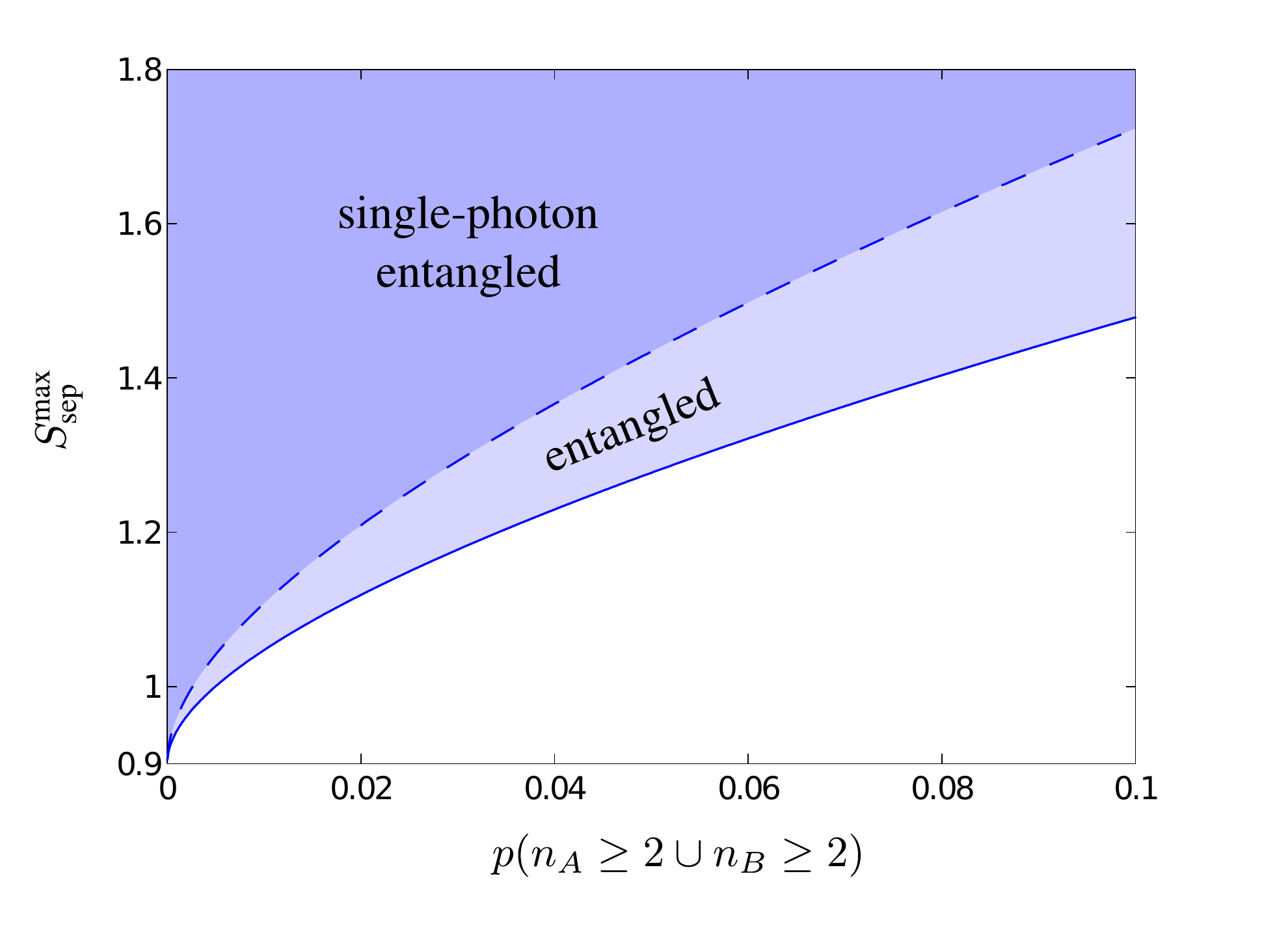}
\caption{Separable bound $S^{\max}_{\rm sep}$ as a function of the probability that at least one of the two protagonists gets more than one photon $p(n_A \geq 2 \cup n_B \geq 2).$ The dashed curve allows one to know the maximum value of the CHSH polynomial $S^{\max}_{\rm sep}$ that a state that is separable in the $\{|0\rangle,|1\rangle\}^{\otimes 2}$ subspace can reach. If the observed CHSH value $S_{\rm obs} > S^{\max}_{\rm sep},$ one can conclude that the projection of the state in the subspace with zero and one photon locally is entangled. The full curve give the maximum value that a separable state can reach. If the the oberved CHSH value is higher than the latter, one can conclude about entanglement but without saying where the entanglement lies.}
\label{fig1_sm}
\end{figure}

Without loss of generality, Alice \& Bob's state can be written as
\begin{equation}
\rho=\begin{pmatrix}
  \rho_{n_A \leq 1 \cap n_B \leq 1}    &  \rho_{\rm{coh}}  \\
    \rho_{\rm{coh}}^\dag  &  \rho_{n_A \geq 2 \cup n_B \geq 2} 
\end{pmatrix}
\end{equation}
where $\rho_{n_A \leq 1 \cap n_B \leq 1}$ denotes the block with at most one photon per mode, $\rho_{n_A \geq 2 \cup n_B \geq 2}$ refers to the block where at least one of the two modes is populated with more than one photon and $\rho_{\rm{coh}}$ is associated to the coherence between these two blocks. Taking the constraints (\ref{constraints1}),(\ref{constraints2}) and (\ref{constraints3}) into account and using the formula (\ref{CHSH_random}), the corresponding CHSH polynomial reduces to
\begin{eqnarray}
\label{S_averaged}
&S&=\frac{16}{\sqrt{2} \pi} \Re\Big[\langle 01 | \rho_{n_A \leq 1 \cap n_B \leq 1} | 10\rangle\Big]\\
\nonumber
&& + \frac{8}{\pi} \left(\Re\Big[\langle 20 | \rho_{\rm{coh}} | 11\rangle\Big]+\Re\Big[\langle 02 | \rho_{\rm{coh}} | 11\rangle\Big]  \right)\\
\nonumber
&& +S_{ \rho_{n_A \geq 2 \cup n_B \geq 2}}
\end{eqnarray}
where $\Re$ denotes the real part and $S_{ \rho_{n_A \geq 2 \cup n_B \geq 2}}$ is associated to the CHSH value obtained from $\rho_{n_A \geq 2 \cup n_B \geq 2}.$ The goal is now to optimize $S$ over the set of separable states. Let $p(n_A \geq 2 \cup n_B \geq 2)$ be the probability that at least one of the two protogonists has more than one photon, i.e.
\begin{equation}
\rm{tr} \-\ \rho_{n_A \geq 2 \cup n_B \geq 2} = p(n_A \geq 2 \cup n_B \geq 2).
\end{equation}
Since $\rho_{n_A \geq 2 \cup n_B \geq 2}$ cannot be obtained by local projections, it may maximally contribute to CHSH, i.e. $S_{ \rho_{n_A \geq 2 \cup n_B \geq 2}} \leq 2\sqrt{2} \times p(n_A \geq 2 \cup n_B \geq 2).$ S is thus upper bounded by
\begin{eqnarray}
\label{tobe_optimized}
&S \leq & S^{\max} (p(n_A \geq 2 \cup n_B \geq 2))\\
\nonumber
&&= \frac{16}{\sqrt{2} \pi} \Re\Big[\langle 01 | \rho_{n_A \leq 1 \cap n_B \leq 1} | 10\rangle\Big]\\
\nonumber
&& + \frac{8}{\pi} \left(\Re\Big[\langle 20 | \rho_{\rm{coh}} | 11\rangle\Big]+\Re\Big[\langle 02 | \rho_{\rm{coh}} | 11\rangle\Big]  \right)\\
\nonumber
&& +2\sqrt{2} \times p(n_A \geq 2 \cup n_B \geq 2).
\end{eqnarray}
Needless to say, separable states are physical states. They are thus represented by positive matrices with a trace (tr) equal to one. Furthermore, the Peres-Horodecki criterion \cite{Peres, Horodecki} states that for any separable state $\rho_{\rm sep},$ its partial transpose $\rho_{\rm sep}^{T_b}$ has non-negative eigenvalues. The optimization of $S^{\max}$ over the set of separable states in the subspace with at most one photon locally is thus a problem that can be summarized as follows
\begin{eqnarray}
\label{optimization}
\max_{\rho \in \{|0\rangle,|1\rangle,|2\rangle\}^{\otimes 2}} &:&  S^{\max}(p(n_A \geq 2 \cup n_B \geq 2)) \\  
 \text{s. t.} &:& \rho \geq 0 \nonumber\\
 &&{\rm tr} (\rho) \leq 1\nonumber\\
 && \left(\prod_{0/1}\rho\prod_{0/1}\right)^{T_b} \geq 0 \nonumber\\
 && \sum_{i,j=0}^{1} p_{ij}=1-p(n_A \geq 2 \cup n_B \geq 2). \nonumber
 \end{eqnarray} 
The constraint ${\rm tr} (\rho) \leq 1$ comes from the fact that the optimization is performed over finite dimension $(9 \times 9)$ matrices that can either represent a physical state or that can be obtained by local projections of states spanning Hilbert spaces with a larger dimension. $\prod_{0/1}$ stands for the projection into the subspace with at most one photon locally. Eq. (\ref{optimization}) is a linear optimization with semidefinite positive constraints which can be efficiently solved numerically \cite{Boyd0}. The result of the optimization $S_{\rm sep}^{\max}$ is shown in Fig. 2 (main text) as a function of $p(n_A \geq 2 \cup n_B \geq 2).$ It is also given in Fig. \ref{fig1_sm} (dashed line). If a physical state that satisfies the last condition leads to a $S$ value larger than $S_{\rm sep}^{\max},$ one concludes that the partial transpose has at least one negative eigenvalue and hence, the state is entangled and entanglement lies in the subspace with zero and one photon locally, i.e. refers to single-photon entanglement.\\

It is worth mentioning that if the observed CHSH value is not higher than the separable bound obtained through the optimization presented in Eqs. (\ref{optimization}), one can still conclude about the presence of entanglement. Indeed, the optimization can be performed over the set of separable states, but not only the ones that are separable in the subspace with at most one photon locally. The constrain $\left(\prod_{0/1}\rho\prod_{0/1}\right)^{T_b} \geq 0$ simply needs to be replaced by $\rho^{T_b} \geq 0.$ This leads to the second separable bound (full curve) presented in Fig. \ref{fig1_sm}. If the observed CHSH value is larger than the bound obtained in this manner, one can conclude about the presence of entanglement but we cannot say in which subspace entanglement lies. \\

Note also that the optimization (\ref{optimization}) does not take the individually measured probabilities $p(n_A=j)$ $(p(n_B=j))$ for having $j$ photons in Alice's (Bob's) mode into account. These probabilities add constraints reducing the size of the set of separable states over which $S^{\max}$ is optimized and thus, provide a tighter bound $S_{\rm sep}^{\max}.$ We use them for the calculation of the separable bounds related to the experiment (see the optimization (\ref{sum_optimization_errors})).  

\subsection{General procedure to follow for witnessing single-photon entanglement}
Below, we present the procedure to follow in order to conclude about the presence of entanglement with the proposed witness.

\begin{itemize}
\item 
Firstly, the local photon number distributions $p(n_a=0),p(n_b=0), p(n_a=1),p(n_b=1)$ are accessed using local phase-averaged quantum state tomography. 
\item The joint probability $p(n_a \geq 2 \cup n_b \geq 2)$ defined by $$p(n_a \geq 2)+ p(n_b \geq 2)-p(n_a \geq 2 \cap n_b \geq 2)$$ is then upper bounded by
\begin{eqnarray}
\nonumber
&p^\star =& 2-p(n_a=0)-p(n_a=1)\\
\nonumber
&&-p(n_b=0)-p(n_b=1).
\end{eqnarray}
\item Thirdly, $p^\star$ is used to obtain the separable bound $S^{\max}_{\rm sep}(p^{\star})$ deduced from the following optimization 
\begin{eqnarray}
\label{sum_optmization_proc}
\max_{\rho \in \{|0\rangle,|1\rangle,|2\rangle\}^{\otimes 2}} &:&  S^{\max}(p^{\star}) \nonumber\\  
 \text{s. t.} &:& \rho \geq 0 \nonumber\\
 &&{\rm tr} (\rho) \leq 1\nonumber\\
 && \left(\prod_{0/1}\rho\prod_{0/1}\right)^{T_b} \geq 0 \nonumber\\
 && \sum_{i,j=0}^{1} p_{ij} \geq 1-p^{\star}.
 \end{eqnarray}
Alternatively, $S^{\max}_{\rm sep}(p^{\star})$ can simply be obtained from Fig. 2 (main text) because $S^{\max}_{\text sep}$ is a monotonically increasing function of $p(n_a \geq 2 \cup n_b \geq 2).$
Both methods provide an upper bound on $S^{\max}_{\text sep}(p(n_a \geq 2 \cup n_b \geq 2)) \leq S^{\max}_{\text sep}(p^{\star}).$
\item Fourthly, the CHSH value is measured. In principle, this is done by proposing Alice and Bob to randomly choose measurements among $\{X_{\varphi_A^1}, X_{\varphi_A^2}\}$ and $\{X_{\varphi_B^1}, X_{\varphi_B^2}\}$ respectively and to subsequently compute the CHSH polynomial. However,  since $E_{X_{\varphi_A^1},X_{\varphi_B^1}}=-E_{X_{\varphi_A^2},X_{\varphi_B^2}}$ and $E_{X_{\varphi_A^1},X_{\varphi_B^2}}=E_{X_{\varphi_A^2},X_{\varphi_B^1}},$ $S_{\text obs}$ can be obtained in practice from the measurements of two correlators only
\begin{equation}
\label{S_obs}
S_{\text obs}=2 E_{X_{\varphi_A^1},X_{\varphi_B^1}}+2 E_{X_{\varphi_A^1},X_{\varphi_B^2}}. 
\end{equation} 
\item Finally, $S_{\text obs}$ and $S^{\max}_{\text sep}(p^{\star})$ are compared. If $S_{\text obs} > S^{\max}_{\text sep}(p^{\star})$ (which implies that $S_{\text obs} > S^{\max}_{\text sep}(p(n_a \geq 2 \cup n_b \geq 2))$), one concludes that the measured state is entangled at the single-photon level. Otherwise, we cannot form a conclusion, the state can either be separable or entangled. \\
\end{itemize}

\section{Appendix B: Experimental details}
\subsection{Heralded creation of single-photon entanglement}
A continuous-wave frequency-doubled Nd:YAG laser (Diabolo, Innolight) pumps a triply-resonant type-II phase-matched optical parametric amplifier based on a KTP crystal to generate below threshold, a two-mode squeezed state \cite{Laurat}. The created modes are orthogonally polarized and are deterministically separated at the output of the amplifier. One of the two modes is send to a single-photon detector (superconducting single-photon detector, with a quantum efficiency of 7$\%$ at 1064nm) after filtering of the non-degenerate modes. A detection event ideally heralds the generation of a single-photon state on the twin mode. The heralding rate is around 30 kHz. (See \cite{Morin} for more details about the source). By controlling the polarization of the heralded photon via a half-wave plate and by subsequently sending it into a polarizing beam splitter (PBS), one gets a versatile source producing states of the form
\begin{equation}
|\psi(\theta)\rangle_{AB} = \cos(2\theta)|0\rangle_A |1\rangle_B + \sin(2\theta)|1\rangle_A|0\rangle_B
\end{equation}
where $\theta$ is the angle of the half-wave plate relative to the axis of the PBS. For $\theta=0^{\circ},$ the created state is separable, whereas for $\theta=22.5^{\circ}$ it becomes maximally entangled. This source is thus particularly well suited to test the proposed entanglement witness.
\subsection{Homodyning}
Each spatial mode is then detected using an independent homodyne detection. The required local oscillators are obtained by impinging a bright beam into the second input of the PBS. The phase of Alice's local oscillator is controlled relative to Bob's one by choosing appropriately the polarization of the bright beam just before the PBS \cite{Laurat}. To meet the witness requirements, a phase averaging is realized by sweeping a piezoelectric transducer located on the path of the bright beam before the PBS. The overall efficiency of each homodyne detection is 85$\%$, including the quantum efficiency of the photodiodes (Fermionics 500), mode overlap and electronic noise \cite{Kumar}. 
\subsection{Data acquisition}
\begin{figure*}[ht!]
\centering
\includegraphics[width=16 cm]{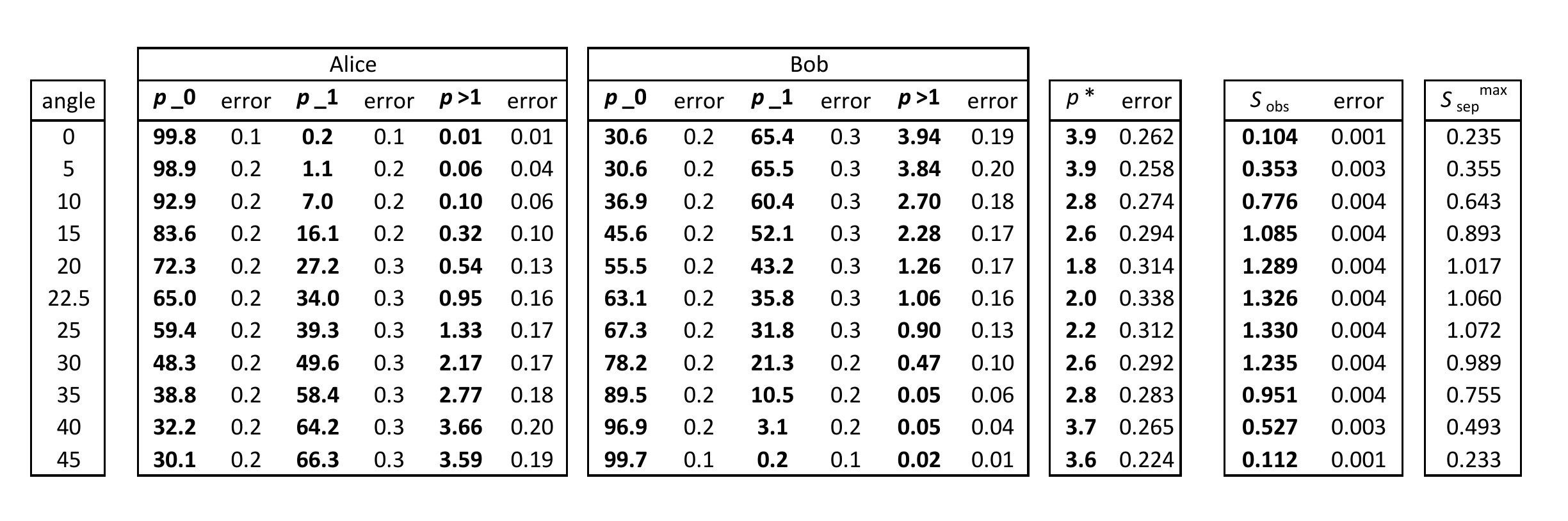}
\caption{Results of local measurements versus the angle of the half-wave plate relative to the axis of the PBS.
The table on the left gives the photon number distribution in Alice's location. ($p\_0,$ $p\_1,$ $p>1$ denote the probability that zero $p(n_A=0),$ one $p(n_A=1)$ and more than one photon $p(n_A \geq 2)$ occupy Alice's mode. The second table gives Bob's results. The two last tables give an upper bound on the probability that at least one of the two protagonists gets more than one photon $p(n_A \geq 2 \cup n_B \geq 2)$ and the observed CHSH value $S_{\text obs}.$)}
\label{fig2_sm}
\end{figure*}

For each heralding event, Alice and Bob perform a quadrature measurement and the corresponding result is extracted from homodyne photocurrents. 200000 events are accumulated for each relative phase $\Delta\varphi_{AB}^{11}=\frac{\pi}{4}$ and $\Delta\varphi_{AB}^{12}=-\frac{\pi}{4}.$ $S_{\text obs}$ is then deduced from Eq. (\ref{S_obs}). The same results are also used to compute the local photon number distributions by phase-averaged quantum state tomography using the method given in \cite{Munro}. We remind that this method does not require assumptions on the the dimension of the measured state. Table \ref{fig2_sm} shows the results for various angles $\theta.$ Alternatively, a MaxLike algorithm  \cite{Lvovsky} can be used to access the local photon number distribution, albeit with a truncation of the Fock space.

\subsection{Error estimations}
Several kind of errors should be taken into account. 

Firstly, statistical errors affect the measured value of the CHSH polynomial $S_{\text obs}$. These errors are estimated in a standard way by using the central limit theorem. They are basically very small (see Table \ref{fig2_sm}) because they were deduced from 200000 results.\\
Secondly, the accuracy with which the relative phase $\Delta\varphi_{AB}^{\ell \bar \ell}$ between Alice \& Bob's measurements is estimated to be $2^\circ.$ This means that when Alice and Bob choose measurement settings, e.g.  corresponding ideally to the quadratures $X_{\varphi_A^{1}}$-$X_{\varphi_B^{1}},$ the relative phase $\Delta\varphi_{AB}^{11}$ is not exactly equal to $\frac{\pi}{4}$ as it should be, but $\Delta\varphi_{AB}^{11}=\frac{\pi}{4}+\epsilon^{11}$ where $-1^{\circ} \leq \epsilon^{11} \leq 1^{\circ}.$ This error is taken into account into the separability bound. \\
Thirdly, errors also affect the local photon-number probability distributions, which are estimated using the phase-averaged homodyne measurements \cite{Munro}. Evaluating these errors is a not trivial task. We use the following method for estimating the overall error. The tomography that we use to access the local photon number distributions leaves us with the diagonal elements of an estimated density matrix $\rho_{\rm estimate}.$ Using this matrix, we simulate the quadrature data, then reconstruct the diagonal elements of a simulated density matrix $\rho_{\rm simul}.$ The approach is repeated 200 times, always from the same initial state $\rho_{\rm estimate}.$ This generates a random set of 200 data points for each local probability $p(n_A=0), p(n_B=0), p(n_A=1), p(n_B=1) \hdots$ and the corresponding standard deviation provides the desired error $\delta p(n_A=0), \delta p(n_B=0), \delta p(n_A=1),  \delta p(n_B=1) \hdots$\\

We now show how the errors on the measurement angles $\epsilon^{\ell \bar \ell}$ and on the local photon-number probability distributions $\delta p(n_A=i), \delta p(n_B=j)$ have been taken into account in the calculation of the separable bound $S_{\text sep}^{\max}.$ Since, in practice, it is necessary to measure two correlators only, the separable bound can be calculated under the assumption that the measurements that have been performed are such that 
\begin{eqnarray}
\nonumber
&&\Delta\varphi_{AB}^{11}=-\frac{\pi}{4} + \epsilon^{11},
\Delta\varphi_{AB}^{22}=\frac{3\pi}{4} + \epsilon^{11},\\
\nonumber
&&\Delta\varphi_{AB}^{12}=\Delta\varphi_{AB}^{21}=\frac{\pi}{4} + \epsilon^{12},
\end{eqnarray}
i.e. that the errors are the same for $\Delta\varphi_{AB}^{11}$ and $\Delta\varphi_{AB}^{22}$ and for $\Delta\varphi_{AB}^{12}$ and $\Delta\varphi_{AB}^{21}.$ These erroneous measurement angles yield to an upper bound $S^{\max}$ (which replaces the one in Eq. (\ref{tobe_optimized})) given by
\begin{eqnarray}
\label{tobe_optimized_angle+prob}
 \nonumber
&S \leq & S^{\max} (p(n_A \geq 2 \cup n_B \geq 2), \epsilon^{11},  \epsilon^{12})\\
\nonumber
&&
= \frac{4}{\pi} \left( \Re \Big[\langle 10 | \rho | 01\rangle \Big] \times C - \Im  \Big[\langle 10 | \rho | 01\rangle\Big] \times D \right)\nonumber \\
&&+ \frac{4}{\sqrt{2} \pi} \Re\Big[\langle 20 | \rho | 11 \rangle + \langle 11 | \rho | 02 \rangle \Big] \times C\nonumber \\
&&- \frac{4}{\sqrt{2} \pi} \Im\Big[ \langle 20 | \rho | 11 \rangle + \langle 11 | \rho | 02 \rangle \Big] \times D \nonumber \\
&&+2\sqrt{2} \times p(n_A \geq 2 \cup n_B \geq 2)
\end{eqnarray}
where $\Im$ denotes the imaginary part and
\begin{eqnarray}
\nonumber
C &=& 2 \left(\cos(\epsilon^{11}-\frac{\pi}{4}) + \cos(\epsilon^{12}+\frac{\pi}{4})\right),\\
\nonumber
D &=& 2 \left(\sin(\epsilon^{11}-\frac{\pi}{4}) + \sin(\epsilon^{12}+\frac{\pi}{4})\right).
\end{eqnarray}
$S^{\max}$ can now be optimized over the set of separable states, as before. If one takes into account the local photon number probability distribution and the errors as well, one ends up with the following optimization problem 
\begin{eqnarray}
\label{sum_optimization_errors}
 max&:&  S^{\max}(p^{\star}, \epsilon^{11}, \epsilon^{12}) \\  
 s.t. &:& \rho \geq 0 \nonumber\\
 &&{\rm tr} (\rho) \leq 1\nonumber\\
 && \left(\prod_{0/1}\rho\prod_{0/1}\right)^{T_b} \geq 0 \nonumber\\
 && -1^{\circ} \leq \epsilon^{11}, \epsilon^{12} \leq +1^{\circ} \nonumber\\
 && p_{00}+p_{10}+p_{01}+p_{11} \geq 1- p^{\star}-\delta p^{\star}  \nonumber\\
 &&  p_{00} + p_{01} + p_{02} \leq p(n_A=0)+\delta p(n_A=0) \nonumber\\
 && p_{10} + p_{11} + p_{12} \leq p(n_A=1)+\delta p(n_A=1) \nonumber\\
 && p_{20} + p_{21} + p_{22} \leq p(n_A > 1)+\delta p(n_A> 1)  \nonumber\\
 && p_{00} + p_{10} + p_{20} \leq p(n_B = 0)+\delta p(n_B=0) \nonumber\\
 && p_{01} + p_{11} + p_{21} \leq p(n_B = 1)+\delta p(n_B=1) \nonumber \\
 &&p_{02} + p_{12} +p_{22} \leq p(n_B>1)+\delta p(n_B>1)\nonumber 
 \end{eqnarray}
In addition to errors, we emphasize that the previous optimization uses the knowledge of each local photon number probability $p(n_A=0),$ $p(n_A=1) \hdots$ This provides a tighter separable bound as compared to the optimization summarized in Eq. (\ref{sum_optmization_proc}).\\
Note that the optimization (\ref{sum_optimization_errors}) is non-linear in $\epsilon^{11}$ and $\epsilon_{12}.$ Hence, the result cannot be obtained with standard semidefinite solvers. However, one can convince oneself that the maximum value of $S_{sep}^{\max}(p^{\star}, \epsilon^{11}, \epsilon^{12})$ is obtained for the extremal choice $\epsilon^{11}=1^{\circ},$ $\epsilon^{12}=-1^{\circ}.$ This can be understood intuitively since this choice brings the settings closer to each other and thus, helps in increasing the CHSH value. Note that the linearity of the optimization is recovered once the values of $\{\epsilon^{11}, \epsilon^{12}\}$ are fixed. The results of the optimization for $\epsilon^{11}=1^{\circ}$ and $\epsilon^{12}=-1^{\circ}$ are shown in Fig. 4 (main text). \\


\begin{thebibliography}{3}
\bibitem{Note_Babichev04} Note that in Ref. \cite{Babichev04}, in addition to a tomographic approach, a Bell test is presented which can be used for witnessing single-photon entanglement. The latter relies on the rejection of some of the acquired data. In comparison, our witness does not need post-selection.
\bibitem{Comment_tomo} The number of measurement that are needed to reconstruct a state of dimension d shared by n parties (nodes) is given by $(d+1)^n=e^{n\log (d+1)}.$
\bibitem{explain} The witness does not require one to know the way in which the state was created nor its dimensionality. However, it requires that Alice \& Bob's devices perform the desired measurements. In particular, for the proposed witness, measurements are assumed to be homodyne measurements defined in the single optical mode of a local oscillator.
\end{thebibliography}
\end{document}